\documentclass[preprint,10pt, a4paper]{elsarticle}
\usepackage[left=2cm,right=2cm,top=2cm,bottom=2cm]{geometry}
\usepackage{eurosym}
\usepackage[pagewise]{lineno}
\usepackage{svg}
\usepackage{soul}
\usepackage[utf8]{inputenc}
\usepackage{amssymb, bm}
\usepackage{pgfplots}
\usepackage{multirow}
\usepackage{amsfonts}
\usepackage{comment}
\usepackage{booktabs}
\usepackage{siunitx}
\usepackage{xcolor}
\usepackage{float}
\usepackage{caption}
\usepackage{subcaption}
\usepackage{newfloat}
\DeclareFloatingEnvironment[name={Listing}]{codefig}

\usepackage[ruled,vlined, noend]{algorithm2e}
\usepackage{caption}
\usepackage{subcaption}
\usepackage{amsmath}
\definecolor{rev}{rgb}{0.0, 0.0, 0.0}

\usepackage{paralist}
\usepackage{enumitem}

\usepackage{tikz}
\usepackage[draft]{todonotes}
\usepackage{amsopn}
\usepackage{mdframed}

\usetikzlibrary{arrows.meta}

\usepackage{amssymb}
\usepackage{lineno}

\usepackage{float}
\restylefloat{table}

\journal{SoftwareX}

\begin{document}

\begin{frontmatter}

%% Title, authors and addresses

%% use the tnoteref command within \title for footnotes;
%% use the tnotetext command for theassociated footnote;
%% use the fnref command within \author or \address for footnotes;
%% use the fntext command for theassociated footnote;
%% use the corref command within \author for corresponding author footnotes;
%% use the cortext command for theassociated footnote;
%% use the ead command for the email address,
%% and the form \ead[url] for the home page:
%% \title{Title\tnoteref{label1}}
%% \tnotetext[label1]{}
%% \author{Name\corref{cor1}\fnref{label2}}
%% \ead{email address}
%% \ead[url]{home page}
%% \fntext[label2]{}
%% \cortext[cor1]{}
%% \address{Address\fnref{label3}}
%% \fntext[label3]{}

\title{MaaSSim - agent-based two-sided mobility platform simulator}

\author[1,*]{Rafal Kucharski}
\author[1]{Oded Cats}

\address[1]{Department of Transport \& Planning, TU Delft, Netherlands}
\address[*]{Corresponing author: r.m.kucharski@tudelft.nl}

\begin{abstract}
%% Text of abstract 

Two-sided mobility platforms, such as Uber and Lyft, widely emerged in the urban mobility landscape, bringing disruptive changes to transportation systems worldwide. 
This calls for a simulation framework where researchers from various and across disciplines may introduce models aimed at representing the dynamics of platform-driven urban mobility systems.

In this work, we present MaaSSim, an agent-based simulator reproducing the transport system used by two kind of agents: (i) travellers, requesting to travel from their origin to destination at a given time, and (ii) drivers supplying their travel needs by offering them rides. An intermediate agent, the platform, allows demand to be matched with supply. Agents are decision makers, specifically, travellers may decide which mode they use or reject an incoming offer. Similarly, drivers may opt-out from the system or reject incoming requests. All of the above behaviours are modelled through user-defined modules, representing agents' taste variations (heterogeneity), their previous experiences (learning) and available information (system control). MaaSSim is an open-source library available at a public repository\footnote{\texttt{github.com/RafalKucharskiPK/MaaSSim}}, along with a set of tutorials and reproducible use-case scenarios.

\end{abstract}


\end{frontmatter}

\section*{Current code version}
\label{}

\begin{table}[H]
\small
\begin{tabular}{|l|p{4cm}|p{11cm}|}
\hline
\textbf{Nr.} & \textbf{Code metadata description} & \textbf{Please fill in this column} \\
\hline
C1 & Current code version & 0.9.5 \\
\hline
C2 & Permanent link to code/repository used for this code version &  https://github.com/RafalKucharskiPK/MaaSSim/releases/tag/0.9.5 \\
\hline
C3 & Code Ocean compute capsule & \\
\hline
C4 & Legal Code License   & MIT License \\
\hline
C5 & Code versioning system used & git \\
\hline
C6 & Software code languages, tools, and services used & python \\
\hline
C7 & Compilation requirements, operating environments \& dependencies & python 3.7+, simpy, networkX, osmnx, pandas \\
\hline
C8 & If available Link to developer documentation/manual &  https://github.com/RafalKucharskiPK/MaaSSim/tree/master/docs/tutorials \\
\hline
C9 & Support email for questions & \texttt{r.m.kucharski at tudelft.nl} \\
\hline
\end{tabular}
\caption{Code metadata (mandatory)}
\label{} 
\end{table}

%\linenumbers

\section{Motivation and significance}
\label{}

Two-sided mobility platforms (like Uber and Lyft) match supply with demand in urban transportation systems. 
Users submit travel requests in real-time and are instantly matched with drivers, offering to take them to their desired destination. 
All parties are independent decision makers acting according to their individual, heterogeneous preferences and learning from past experiences. 
Travellers are free to select among competing platforms and travel modes, whereas drivers choose whether to work for the available platforms, their working hours and strategically select served requests to maximise revenues. 

%Mobility platforms operate on a dense urban environment, with its own complex topology and traffic congestion conditions varying in time and space. 
%Moreover, platform operations may contribute to adverse road network conditions, for example when they induce demand for new trips or shifts from public transport and active modes. The emerging dynamics are not trivial and often yield feedback loops.

Understanding and modelling such systems is challenging and requires a broad set of expertise. 
While modelling the relations between the different actors involved in two-sided platforms proved to be challenging \cite{casey2012dynamics, graham2017digital}, their manifestation in the context of a dense and congested urban mobility networks induces additional complexity. 
Some of the prevalent topics addressed by researchers include:
supply-demand interactions \cite{xu2020supply, nourinejad2019ride};
optimal matching of drivers to requests \cite{cui2020understanding, zuniga2020evaluation, suhr2019two};
travellers' mode and platform choices \cite{welch2020shared, wang2020empirical, alonso2020drivers, alonso2020value};
drivers' participation, working shifts and platform choices \cite{ashkrof2020understanding, bokanyi2020understanding}; 
labour economics impact \cite{berger2018drivers, fielbaum2020sharing, allon2018impact}; 
platform pricing strategies \cite{ke2020pricing}; 
pooled rides \cite{kucharski2020exact, Alonso-mora2018, maciejewski, veve2020estimation}; 
impact of shared autonomous vehicles \cite{fagnant2018dynamic, engelhardt2019quantifying, horl2019fleet}; 
fleet size determination \cite{vazifeh2018addressing}
and driver re-positioning strategies \cite{repos,afeche2018ride}.

Each of these research domains involves a series of significant and challenging research questions. Answering each of which is non-trivial, yet the main challenge lays in representing the complete system with its (inter)dependencies and feedback loops, non-determinism and adaptive evolution. In the absence of an encompassing modelling framework, most studies have been limited to a single aspect, e.g. \cite{kucharski2020exact} focuses on travellers behaviour and neglects fleet operations, whereas \cite{Alonso-mora2018} focuses on real-time fleet operations while neglecting the travellers decision process; \cite{maciejewski} embeds shared taxi operations into a detailed traffic agent-based micro-simulation, while \cite{bokanyi2020understanding} relies on arbitrary grid network and focuses on income equity. 

To this end, we provide a modular, extensible framework which contains the fundamental representation of key phenomena related to two-sided mobility platforms.
The key contribution of our software is in advancing the state-of-the-art of agent-based transport models towards two-sided mobility platform systems by providing a framework for developing, testing and simulating platform operations under various settings. 
An extensive set of tutorials and sample experiments facilitates a fast learning curve for users of various background, while modular, extensible architecture allows for its seamless development.

\section{Software description}
MaaSSim is an agent-based simulator, reproducing the dynamics of two-sided mobility platforms in the context of urban transport networks. It models the behaviour and interactions of two kind of agents: (i) travellers, requesting to travel from their origin to destination at a given time, and (ii) drivers supplying their travel needs by offering them rides. The interactions between the two agent types are mediated by the platform, matching demand and supply. Both supply and demand are microscopic. For supply this pertains to the explicit representation of single vehicles and their movements in time and space (using a detailed road network graph), while for demand this pertains to exact trip request time and destinations defined at the graph node level. Agents are decision makers. Specifically, travellers may reject the incoming offer or decide to use another mode than those offered by the mobility platform altogether. Similarly, driver may opt-out of the system (stop providing service) or reject/accept incoming requests. Moreover, they may strategically re-position while being idle. 

All of above behaviours are modelled through user-defined \textbf{decision modules}, by default deterministic, optionally probabilistic, representing agents' taste variations (heterogeneity), their previous experiences (learning) and available information (system control). 
%Similarly, the system performance (amongst others travel times and service times) may be deterministic or probabilistic. 
Each simulation run results in two sets of outputs. The sequence of recorded space-time locations and statuses for simulated vehicles and travellers. These outputs are further synthesised into agent-level and system-wide KPIs for in-depth analyses.
 
\subsection{Software Architecture}
\label{}

\begin{figure}
    \centering
    \includegraphics[width=8cm]{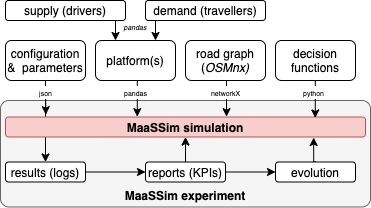}
    \caption{Input and output of MaaSSim workflow}
    \label{fig:1}
\end{figure}

\begin{figure}
    \centering
    \includegraphics[width=\columnwidth]{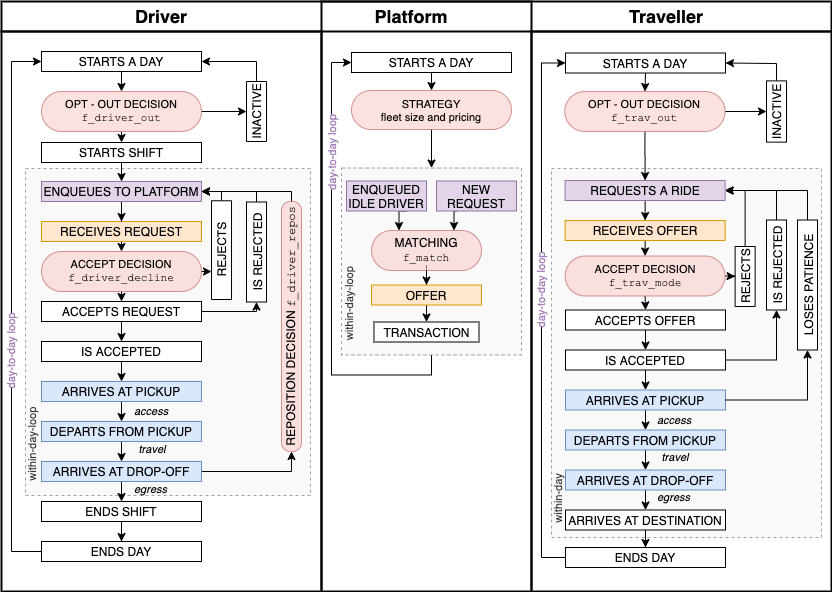}
    \caption{Routines of the three kinds of agents in MaaSSim. Boxes in violet denote an interaction with the platform, orange displays matching between drivers and travellers, blue refers to joint part where a traveller is transported by the driver. Places where agents make a decision are marked with red rounded boxes and their decision protocols can be replaced by used-defined python functions.}
    \label{fig:2}
\end{figure}

MaaSSim is a scalable, easy-to-use, modular, extensible python library.
The main class is called with a configuration file (\texttt{.json} file) allowing to control the input (travel demand, fleet supply, road network) and simulation (e.g. event duration and their variability). External decision functions to reproduce desired agents' behaviour are passed by reference and can be user-defined (fig. \ref{fig:1}).
A simulation corresponds to a single day, during which routines of interacting agents are processed (fig. \ref{fig:2}) with a $Simpy$ discrete-event simulation framework \cite{simpy}.
%Typical simulation consists of multiple requests, travellers and platforms and is simulated over a long time period (e.g. 10000 travellers served by 1000 drivers assigned to one of 3 platforms simulated over 8 hours on the Amsterdam network of 50 000 nodes). 
The simulation outputs raw logs (where spatio-temporal stamps of consecutive events are stored for each agent) as well as aggregated reports. 

MaaSSim allows to replicate simulations (to obtain meaningful distributions of random variables), explore multidimensional parameter grids in parallel (e.g. various travellers' value-of-time and fleet size combinations) and simulate day-to-day evolution until convergence.
Independent simulation runs may be executed in parallel, distributing computation load over multiple threads. 
Listing 1 provides an overview of MaaSSim usage.

 \begin{codefig}[H]
     \centering
     \includegraphics[width=\columnwidth]{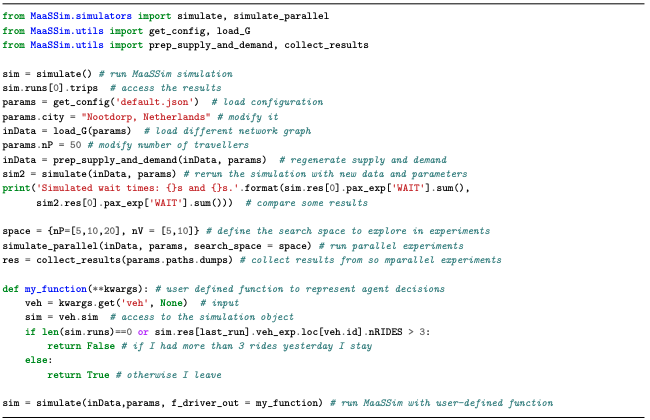}
     \caption{\small MaaSSim usage at glance: starting from a single simulation, through modifying input and configuration, up to parallel experiment computation and user defined decision function.}
     \label{fig:3}
 \end{codefig}

\subsubsection{Input} 
MaaSSim is controlled via $.json$ file which configures the simulation and specifies input. Running the MaaSSim simulation requires: a) an urban road network graph (an instance of \texttt{networkX DiGraph} imported for the simulated urban area with \texttt{OSMnx} \cite{boeing2017osmnx}), b) travel demand (\texttt{pandas DataFrame} with origin and destination nodes and departure time for each trip request) and c) supply specifications (drivers with their working shifts and initial locations). Both supply and demand may come from external data sources or be generated using internal MaaSSim procedures. 

\subsubsection{Agents}
Three kind of MaaSSim agents (implemented as \texttt{SimPy} processes) interact with each other during the course of their daily routines (fig. \ref{fig:2}):
\begin{itemize}
\item \textbf{Travellers} may be assigned to multiple platforms and submit request to all of them to choose the best offer amongst those. A traveller unsatisfied with previous experience may opt-out before requesting. When receiving an offer the traveller makes a decision whether to accept it or not. While accepting she/he walks to the pick-up point, waits until the driver arrives, travels to the drop-off point and walks to the final destination, which terminates traveller's daily routine. 

\item \textbf{Drivers} operate in a loop, queuing to the platform and serving matched requests until the end of their shift (fig. \ref{fig:2}). Along their routines, drivers decide whether to: opt out before starting a shift and not enter the platform at all; accept or reject the incoming requests and finally re-position after becoming idle. 

\item \textbf{Platforms} operate in an infinite loop during simulation. 
Whenever a trip is requested or a driver becomes idle (starts a shift or completes previous request), the platform matches a two-sided queue of travellers on one side and drivers on another. 
A generic algorithm, matching travellers to the closest idle driver, can be replaced by referencing to a user-defined function.

\end{itemize}
\subsubsection{Decision modules} 
The central functionality of MaaSSim is representing agents' individual decision processes (marked with round boxes on fig. \ref{fig:2}). To this end, we introduced an interface where default functions may be overwritten with user-defined modules and integrated within MaaSSim simulations. User-defined functions can use MaaSSim parameterization, internal objects and their attributes (e.g. individual preferences of agents) to reproduce the desired behaviour of:
\begin{itemize}
    \item drivers leaving the system (\texttt{f\_driver\_out}), accepting requests (\texttt{f\_driver\_decline}) and re-positioning (\texttt{f\_driver\_repos}),
    \item travellers leaving the system (\texttt{f\_trav\_out}) and selecting among platforms and modes (\texttt{f\_trav\_mode}),
    \item platform matching requests to drivers (\texttt{f\_match}).
\end{itemize}

% \paragraph{Congestion} is by default static and invariant to MaaSSim results. To focus on phenomena closely related to two-sided operations we assume travel time is exogenous variable, reproduced through fixed network-speed, such approach allows fast computation times. Optionally, travel times may become a random variables following a user-defined distribution. Moreover, to reproduce congestion equilibrium-lile effects, travel times may be updated after each simulation during day-to-day experiment, until convergence (e.g. using MFD approximations of \cite{beojone2020inefficiency})

\subsubsection{Computation times}
MaaSSim has been developed to facilitate the research and assessment of system operations rather than real-time applications. 
The focus of its development has therefore been on code clarity and its accessibility for a broad community. 
Notwithstanding, it remains efficient and allows to run real-size computations within reasonable times. 
For instance, simulating 1000 travellers over 4 hours for the city of Delft, Netherlands takes ca 70s. 
It requires 28 minutes to simulate the Amsterdam network for 8 hours with 5 000 travellers, whereas simulating 50 travellers in the small Nootdorp takes less than 2 seconds on MacBookPro 2019. 
Parallel computations on multiple threads allowed to run 2000 simulations of experiment from fig.\ref{fig:E2} in less than two hours.

The complexity grows with the number of travellers, drivers and platforms (while each agent adds a new routine, the number of possible interactions between travellers and drivers within the platform follows a quadratic pattern). Surprisingly, network size  does not affect the computations, as long as the pre-computed distance matrix fits into memory. 
The computation times may of course increase significantly if complex decision modules are introduced (and executed along with each agent's routine).

\section{Illustrative Examples}
\label{}

We illustrate MaaSSim modelling capabilities through a series of experiments where we simulate various system settings under a range of configurations.

We start with a single simulation, where 10 drivers serve 200 trip requests in Delft, the Netherlands. A single platform matches travellers to the closest drivers who meet at the pick-up point and travel together to the destination. We replicate non-deterministic trip request generation (origin, destination and time) to obtain meaningful results. We analyse them in terms of waiting times, a key performance indicator for both supply and demand. Both agent groups aim to minimise their waiting times, travellers to arrive faster at their destination and drivers to minimise their idle times and thus increase revenues. However the obtained spatial patterns reveal conflicting trends (fig. \ref{fig:E1}). While traveller waiting times (left) are low in the Eastern parts and high in the Western parts of Delft, for drivers the opposite trend prevails (right). This reveals interesting interactions between agents and potentially conflicting interests.

\begin{figure}
    \centering
    \includegraphics[width=0.6\columnwidth]{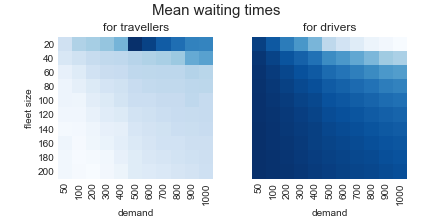}
    \caption{Service performance for various demand and supply levels. Average waiting times for travellers (left) and drivers (right), which follow mirroring diagonal trends. 
    %System performance for traveller improves with increasing supply on one hand and decreasing demand on another, as travellers are served with lower waiting times. Conversely, if demand increases and fleet size decreases, the system improves for drivers, who need to wait less before receiving a request. Yielding an interesting competitive structure, specific to two-sided platforms.
    }
    \label{fig:E2}
\end{figure}

Next, we examine the supply and demand interactions in various settings. We explore the scenario grid varying from low to high demand and supply levels (fig. \ref{fig:E2}). The waiting time for travellers (left) is low when there are few travellers and many drivers. Conversely, drivers idle times decrease if fleet size is low (right). While this overall trend is expected, the magnitude and the sensitivity of these relations and their potential to result with feedback loops on both demand and supply sides of the two-sided market would not have been possible without their detailed modelling. 

\begin{figure}
    \centering
    \includegraphics[width=0.6\columnwidth]{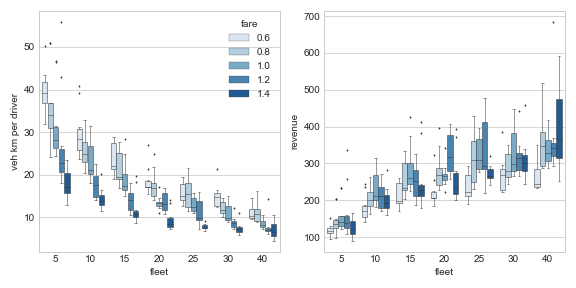}
    \caption{Searching for optimal platform competition strategy: a platform enters a market with competitor operating a fleet of 20 vehicles and offering a trip fare of 1.0 unit/km. We report average vehicle kilometers per driver (left) and total platform revenues (right) resulting from varying fleet size (x-axis) and fare (per-kilometer) combinations for 20 replications.}
    \label{fig:E3}
\end{figure}

Such interaction between the supply and demand typically leads to questions about strategic behaviour, reinforced learning and system equilibria. 
We illustrate how MaaSSim supports answering those questions by means of a platform competition experiment. 
We consider a system where an existing platform with 20 drivers offers travellers a trip fare of 1.0 unit/km. 
We explore potential strategies for a new platform entering the market in terms of two key variables: fleet size (varying from 5 to 40 drivers) and fare (varying from 0.6 to 1.5 units/km). Figure \ref{fig:E3} shows the mileage per driver (left) and total platform revenues (right) resulting from the combination of various strategies. The box plots denote means and distributions resulting from 20 replications. 

Since in the above experiments the simulations are independent (unlike the reinforced learning, where agent decisions are dependent on their experience learned from previous simulation runs) we use parallel computations on multiple threads and collect results in a single file for analysis. 

\begin{figure}
    \centering
    \includegraphics[width=0.4\columnwidth]{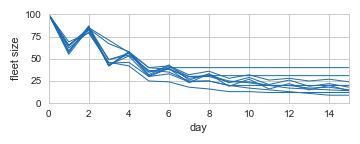}
    \caption{Fleet size evolution for 10 non-deterministic replications of driver reinforced learning behaviour. Drivers make daily decisions to opt out or stay in the system based on previous experience and expected outcomes. %The initial high supply does not allow them to reach the desired income level, so many drivers opt out, yet as the fleet size decreases, the incomes for the remaining drivers increase, making it attractive for drivers to return to the system. Depending on user-defined configuration of the learning process a realistic market evolution may be reproduced
    }
    \label{fig:E4}
\end{figure}

We illustrate the strategic learning of agents with a scenario where 100 drivers serve 200 travellers in a sequence of day-to-day simulations (fig. \ref{fig:E4}). Apparently, the initial supply level is too high, resulting in short waiting times for travellers and low revenues for drivers. Unsatisfied drivers will opt out due to bad previous experiences (low income), adjusting the pool of drivers which decreases until, eventually, some drivers decide to return to the system (as they observe high revenues in the adjusted system). Such adaptation leads to system stabilisation, which may vary due to non-deterministic simulation settings (demand distributions and initial vehicle positions).

\begin{figure}
         \centering
         \includegraphics[width=0.3\columnwidth]{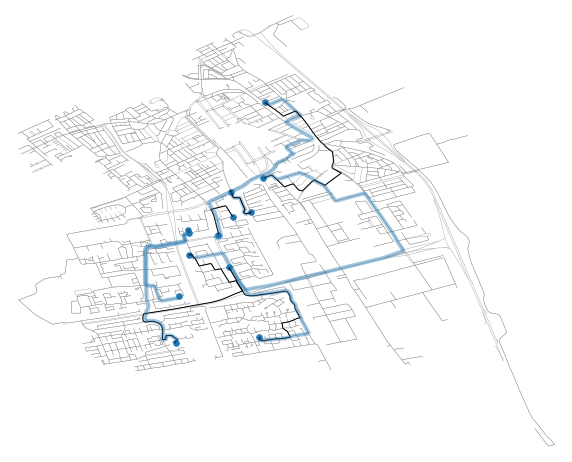}\includegraphics[width=0.3\columnwidth]{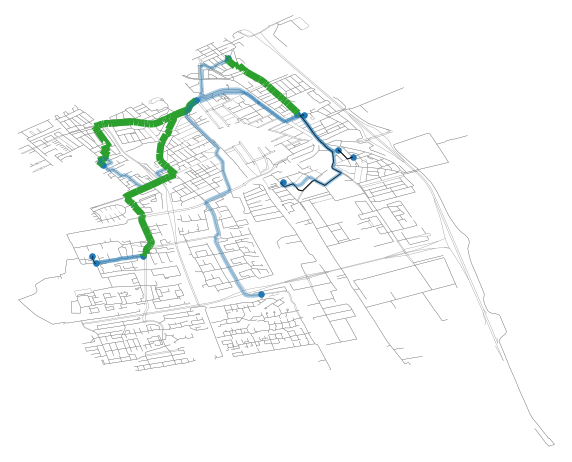}
 
    \caption{Trace of rides for a single simulated vehicle without (left) and with pooled ride services (right). Blue - single traveller on-board, Green - several travellers sharing a ride, Black - empty vehicle trip.}
    \label{fig:E5}
\end{figure}

Finally, we demonstrate how ride-pooling is embedded with simulations on fig. \ref{fig:E5}, where we show a trajectory of a vehicle serving non-shared, private rides (left) and pooled-rides (right). Shared rides are here pre-computed with external ExMAS \cite{kucharski2020exact} algorithm relying on behavioural and system parameters to optimally match travellers into attractive pooled rides.

\begin{figure}
         \centering
         \includegraphics[width=0.3\columnwidth]{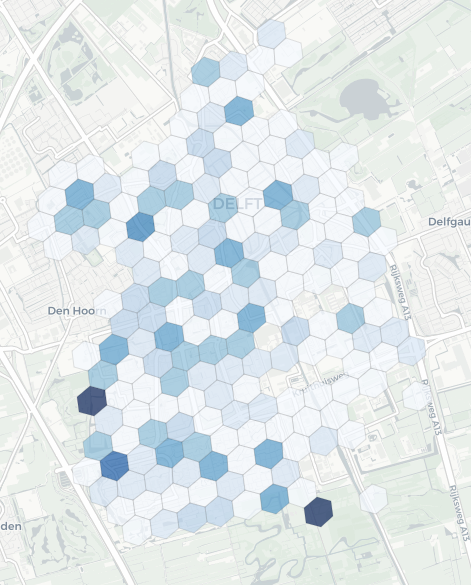}
         \includegraphics[width=0.3\columnwidth]{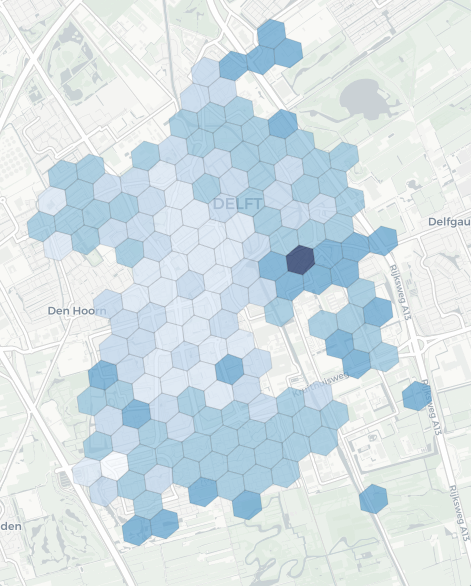}
\caption{Average waiting time for travellers until the driver arrives (left) and for driver, until they get requested (right). Dark denotes longer waiting times. Results from 20 replications of a four hour period simulation with 200 travellers and 10 vehicles in Delft, the Netherlands. 
%While travellers need to wait longer in the western parts of the city, drivers experience short waiting times before getting matched there and their waiting times are longer in the eastern parts, where, in turn, traveller waiting times are shorter.
}
\end{figure}
        \label{fig:E1}

\section{Impact}
\label{}
The emergence of two-sided mobility markets disrupts the transport landscape. Conventional models for transport planning and operations are focused on top-down planning of service lines, timetable and traffic control measures which are not directly suitable for capturing the double-sided dynamics of mobility on demand services. This calls for the development of models that explicitly account for both supply-side and demand-side dynamics, as well as their interaction with the intermediate matching platform. In particular, in order to capture the bottom-up emerging order resulting from two-sided mobility, it is essential to revise the modelling approach of key elements of the transport system: demand (which is now inherently microscopic), supply (which has become a decision maker) and a road network (which capacity and congestion are no longer a single pivot variable of assignment models) along with a new agent, the platform, which orchestrates supply and demand interaction and which might be subject to regulation. For example, New York City has introduced regulations pertaining to guaranteed minimum wage for ride-hailing drivers, entry caps and maximum empty vehicle-kilometre requirements \cite{li2019regulating}. There is therefore a need for a simulation model that facilitates the design and assessment of such policies as well as business models, prior to their implementation. 

A wide array of research questions has consequently emerged, ranging from traffic flow, labour economics, real-time control, optimisation to travel behaviour. This bursting stream of research calls for the development of a unifying simulation framework under which emerging models, algorithms and approaches may be integrated. Recent changes are disruptive enough to justify a new framework, designed to reproduce the dynamics associated with a two-sided platform. 

Furthermore, the already interdisciplinary field of transportation science has recently gained increasing interest from various fields such as complex network theory, system dynamics, social networks, marketing economics and computational physics. This makes it particularly timely to support a fast learning-curve by offering a quick, minimal setup to reproduce the basic dynamics of two-sided mobility platforms. 
To allow researchers to contribute in their domain, the software has to be modular and require minimal knowledge on other modules while allowing to enrich the overall experimental analysis.

MaaSSim is not intended for the complete modelling of transport systems for which there is abundance of mature and developed frameworks, both commercial (like PTV Visum, CUBE, Emme) and open source (like MatSim \cite{w2016multi}, SUMO \cite{SUMO2018}, DynaMIT \cite{ben1998dynamit}, SimMobility\cite{adnan2016simmobility} etc.). Instead, the objective of MaaSSim is to support researches with modelling and reproducing the emerging phenomena taking place in the context of two-sided mobility platform and analyse their disruptive potential for urban transport systems.

\section{Conclusion}
\label{}
The overarching objective underlying the development of MaaSSim is to allow researchers to focus on their partial models and integrate them within the simulation framework, allowing a group of interdisciplinary researchers to share expertise from their fields. For instance, MaaSSim has been instrumental in the scientific discovery process of supply-side dynamics and the fleet size attained in equilibrium in a decentralised bottom-up mobility platform context. This research resulted with a novel service supplier (i.e. driver) learning and choice model \cite{arjen} which can be now reused for studying platform pricing strategies, re-positioning algorithms or travellers mode-choices.

MaaSSim provides an extensible, easy-to-use simulation platform allowing for user-defined representation of two-sided mobility platform that can support a variety of research interests.
MaaSSim is an open-source library available through the \texttt{pip} installer as well as from public repository, where it comes along with a set of tutorials and applicable use-cases. 
With the above set of experiments, coupled with reproducible \texttt{jupyter notebooks} stored on repository, the reader can get an impression of the range of MaaSSim applications and start developing their own experiments. In particular, explore own networks (queried from OpenStreetMap with \cite{boeing2017osmnx}) and run the experiments with their own tailored configuration.

\section*{Conflict of Interest}
We confirm that there are no known conflicts of interest associated with this publication and there has been no significant financial support for this work that could have influenced its outcome.

\section*{Acknowledgements}
This research was supported by the CriticalMaaS project (grant number 804469), which is financed by the European Research Council and Amsterdam Institute for Advanced Metropolitan Solutions.

%% The Appendices part is started with the command \appendix;
%% appendix sections are then done as normal sections
%% \appendix

%% \section{}
%% \label{}

%% References:
%% If you have bibdatabase file and want bibtex to generate the
%% bibitems, please use
%%

\bibliographystyle{elsarticle-num} 
\bibliography{references.bib}

%% else use the following coding to input the bibitems directly in the
%% TeX file.

% \begin{thebibliography}{00}

% %% \bibitem{label}
% %% Text of bibliographic item

% %\bibitem{}

% \end{thebibliography}

\section*{Current executable software version}
\label{}

\begin{table}[H]
\begin{tabular}{|l|p{3.5cm}|p{9.5cm}|}
\hline
\textbf{Nr.} & \textbf{(Executable) software metadata description} & \textbf{Please fill in this column} \\
\hline
S1 & Current software version & 0.9.5 \\
\hline
S2 & Permanent link to executables of this version  & https://github.com/RafalKucharskiPK/MaaSSim/releases/tag/0.9.5 \\
\hline
S3 & Legal Software License & MIT License \\
\hline
S4 & Computing platforms/Operating Systems & iOS, Linux, Microsoft Windows \\
\hline
S5 & Installation requirements & Python 3.7+ \\
\hline
S6 & If available, link to user manual - if formally published include a reference to the publication in the reference list &  https://github.com/RafalKucharskiPK/MaaSSim/tree/master/docs/tutorials \\
\hline
S7 & Support email for questions & \texttt{r.m.kucharski at tudelft.nl}\\
\hline
\end{tabular}
\caption{Software metadata (optional)}
\label{} 
\end{table}

\end{document}